\documentclass[aps,pra,groupaddress,twocolumn]{revtex4-1}
\usepackage{units}
\usepackage{amsmath}
\usepackage{amssymb}
\usepackage{graphicx}
\usepackage{bm}
\usepackage{multirow,color,relsize,ulem,microtype}

\newcommand{\be}{\begin{equation}}
\newcommand{\ee}{\end{equation}}
\newcommand{\ba}{\begin{array}}
\newcommand{\ea}{\end{array}}
\newcommand{\bea}{\begin{eqnarray}}
\newcommand{\eea}{\end{eqnarray}}
\newcommand{\e}{\varepsilon}

\newcommand{\im}[1]{\text{Im}[#1]}

\newcommand{\bx}{\bm{x}}

\begin{document}

\title{Constructing the scattering matrix for optical microcavities as a nonlocal \\boundary value problem}

\author{Li Ge}
\email{li.ge@csi.cuny.edu}
\affiliation{\textls[-18]{Department of Engineering Science and Physics, College of Staten Island, CUNY, Staten Island, NY 10314, USA}}
\affiliation{The Graduate Center, CUNY, New York, NY 10016, USA}

\date{\today}

\begin{abstract}
We develop a numerical scheme to construct the scattering ($S$) matrix for optical microcavities, including the special cases with parity-time and other non-Hermitian symmetries. This scheme incorporates the explicit form of a nonlocal boundary condition, with the incident light represented by an inhomogeneous term. This approach resolves the artifact of a discontinuous normal derivative typically found in the $\cal R$-matrix method.
In addition, we show that by excluding the aforementioned inhomogeneous term, the non-Hermitian Hamiltonian in our approach also determines the Periels-Kapur states, and it constitutes an alternative approach to derive the standard $\cal R$-matrix result in this basis. Therefore, our scheme provides a convenient framework to explore the benefits of both approaches.
We illustrate this boundary value problem using one-dimensional and two-dimensional scalar Helmholtz equations. The eigenvalues and poles of the $S$ matrix calculated using our approach show good agreement with results obtained by other means.\\



\end{abstract}

\date{\today}

\maketitle

\section{Introduction}

Driven by advances in nanofabrication capabilities and their applications to integrated optics, understanding resonances and wave transport in optical microcavities \cite{RKChang,Vahala} has been one of the most energized subjects in modern optics. These compact optical structures also offer a unique opportunity to study non-Hermitian phenomena \cite{RMP} and wave chaos \cite{Nockel,Gmachl} in a well-controlled manner. To probe these properties of optical microcavities, one approach resorts to the scattering matrix formalism \cite{Fan}, which was an essential tool in the understanding of resonances in nuclear physics \cite{Wheeler,Wigner}, particle physics \cite{Nagashima} and quantum field theory \cite{Peskin}, which also played a crucial role in the study of wave transport in various fields including condensed matter systems \cite{Beenakker}, optics \cite{Newton} and microwave networks \cite{Dicke}.

In essence, the scattering matrix, denoted by an energy- or frequency-dependent $S(\omega)$, connects a set of incoming channels $\Psi^-$ to their corresponding outgoing channels $\Psi^+$, both defined outside the scattering potential. Therefore, it takes the openness of the system into account, and the conservation of optical flux in the absence of gain and loss is manifested by the unitarity of $S(\omega)$ (i.e., $S(\omega)S^\dagger(\omega)=\mathbf{1}$). When $S(\omega)$ is analytically continued into the complex-$\omega$ plane, its poles (i.e., where its eigenvalues approach infinity) correspond to the resonances of the system, whose wave functions only connect to the outgoing channels, now also evaluated at complex frequencies \cite{Gamow}.

As already pointed out by Wigner and Eisenbud's early work in nuclear physics \cite{Wigner}, the calculation of the $S$ matrix can be understood as a nonlocal boundary value problem (BVP), which was derived using an orthogonal basis of the system and contains explicitly the real-valued frequencies of this basis. More specifically, this orthogonal basis was defined with vanishing normal derivatives at the boundary of the system, and as a result, the expansion of an arbitrary state $\Psi$ using a finite number of these basis functions has a discontinuous normal derivative in general as an artifact \cite{Wigner,Lane}. Alternative, the expansion can be carried out using quasinormal modes \cite{Breit} (i.e., the Gamow states \cite{Gamow}) or the Periels-Kapur states \cite{Kapur,pra06}, both defined with purely outgoing boundary conditions. These approaches, however, do not remove the artifact in the normal derivative, due to the lack of incoming flux that is inherent in the scattering process. We note that in literature the modal expansion approach, regardless of the specific basis, is referred to as the $\cal R$-matrix method \cite{Lane} in general.

Although the consequence of the aforementioned artifact may be insignificant with the introduction of the Bloch operator \cite{BlochC} and a large number of basis functions \cite{Szmy},
having an approach at one's disposal that addresses this conceptual problem may prove valuable in some cases.
In addition, it is often favorable in optical systems to adopt a self-contained approach, i.e., without requiring \textit{a priori} knowledge or assigning phenomenological parameters to a large set of basis functions.
These two goals can be met in principle by using either time-dependent numerical simulations of the Maxwell's equations \cite{YangP,Shibata} or boundary integral equations \cite{BEM,SCUFFEM}. However, the former do not offer much physical insight on the properties of the $S$ matrix, and the latter require that the Green's function is known inside the optical microcavity and hence do not apply, for example, when the refractive index varies smoothly due to a localized thermal source \cite{Liew} or a modulated optical gain and loss profile \cite{Makris_prl08}.

To overcome these limitations, we propose a finite-difference approach in this article that solves for the steady state of the Maxwell's equations inside a last scattering surface (LSS), outside which the scattered flux does not reenter other parts of the system. As we will show in Sec.~I, accomplishing the two aforementioned goals in one dimension (1D) is effortless, because the boundary conditions are local and involve only the wave function and its spatial derivative at either end of the system. In higher dimensions however, a local boundary condition in the form $f[\Psi(\bx,k), \nabla\Psi(\bx,k)]=0$ that holds for all $\bx$'s on the LSS does not exist in general. Instead, we use a nonlocal boundary condition $f[\Psi(\bx,k), \int_\text{LSS} d\bx' O(\bx,\bx')\Psi(\bx',k)]=0$ in its finite difference form, resulting in (1) a self-contained scheme to construct the $S$ matrix without using a modal expansion that (2) resolves the artifact in the normal derivative at the LSS. In addition, this approach via a nonlocal BVP produces the \textit{same} non-Hermitian Hamiltonian that determines the Periels-Kapur states \cite{Kapur} (or constant-flux (CF) states \cite{pra06} for the Maxwell's equations specifically), which constitutes an alternative approach to derive the standard $\cal R$-matrix result in this basis. Hence our scheme provides a convenient framework to explore the benefits of both approaches when constructing and comprehending the $S$ matrix.

This paper is organized as follows: In Sec. II we discuss the BVP and $\cal R$-matrix approaches in 1D using the scalar Helmholtz equation. Despite the simplicity of the boundary conditions, the discussion already reveals the fundamental connection between the $S$ matrix and the CF states in our approach. We also show explicitly that the (normal) derivative of an arbitrary scattering state cannot be captured accurately by the $\cal R$-matrix approach with a finite number of basis functions. In Sec. III we exemplify the nonlocal BVP approach for the scattering of transverse-magnetic (TM) waves in two dimensions (2D), and the treatment of transverse-electric (TE) waves is very similar. We then apply this scheme to parity-time ($\cal PT$) \cite{Bender1,Bender2,Bender3,El-Ganainy_OL07,Moiseyev,Musslimani_prl08,Makris_prl08,Guo,Longhi,CPALaser,conservation,Robin,EP9,Ruter,Lin,EP9,Feng_NM,Feng2,Walk,Hodaei,Yang,Chang}
and rotation-time ($\cal RT$) symmetric \cite{EP9,conser2D} optical microcavities, focusing on the spontaneous symmetry breaking of the $S$-matrix eigenvalues. Finally, we give some concluding remarks in Sec. IV.

\section{BVP in 1D}
\label{sec:1D}
We begin by considering the 1D Helmholtz equation
\be
[\partial_x^2 + \e(x)k^2]\Psi(x, k) = 0, \label{eq:Helmholtz1D}
\ee
where $\Psi(x,k)$ is the electric field, $\e(x)$ is the dielectric constant for an optical microcavity placed between $-L/2$ and $L/2]$, and $k$ is the free-space wave vector. For an incoming wave from the left (denoted by $\Psi^-_L(x,k) \equiv \exp[ik(x+L/2)]$, we write the formal solution of $\Psi(x,k)$ as
\be
\Psi(x,k) =
\left\{
\begin{array}{ll}
\Psi^-_L + r_L\Psi^+_L,\; &(x<-L/2)\\
t_L\Psi^+_R,&(x>-L/2)
\end{array}
\right.\label{eq:psi_scat}
\ee
where $\Psi^+_{L,R}\equiv\exp(\mp ikx - L/2)$ are the outgoing wave functions on the left and right of the system and $r,t$ are the reflection and transmission coefficients. By requiring that both $\Psi(x,k)$ and $\partial_x\Psi(x,k)$ are continuous at $x=\mp L/2$ (denoted by $\ldots|_{L,R}$), the boundary conditions are then simply
\begin{align}
\partial_x\Psi|_L &= ik(2-\Psi|_L),\label{eq:bd1}\\
\partial_x\Psi|_R &= ik\Psi|_R \label{eq:bd2}
\end{align}
by eliminating $r_L$ and $t_L$, which are \textit{local} without involving both $\partial_x\Psi|_{L,R}$ or $\Psi|_{L,R}$ in a single expression.

Before we embark on our quest to higher dimensions, we note an important feature of Eq.~(\ref{eq:bd1}): without the constant term $2ik$, the boundary conditions become the same as those imposed by the CF states \cite{pra06}, i.e., with purely outgoing waves. Similarly, an incoming wave $\Psi^-_R\equiv \exp[-ik(x-L/2)]$ from the right simply adds an additional constant term $-2ik$ in Eq.~(\ref{eq:bd2}). Therefore, starting from the non-Hermitian Hamiltonian that determines the (outgoing) CF states in the interior of the system, one can obtain the wave function in the scattering problem by turning an eigenvalue problem to an inhomogenous equation. Below we give the specific forms of this non-Hermitian Hamiltonian $H$ and the inhomogenous term $F$ using the finite-difference method.

To start, we discretize the 1D space into $N+2$ equally spaced points, with the left (right) boundary of the optical microcavity placed at the middle of the 0th and 1st ($N$th and $(N+1)$th) points. The separation of two neighboring grid points is then given by $\Delta=L/N$, and the Helmholtz equation takes the following form
\be
\frac{1}{\Delta^2}[\Psi_{i+1} - 2\Psi_{i} + \Psi_{i-1}] + \e_ik^2\Psi_i=0,
\ee
where $\Psi_i,\e_i\,(i=1,2,$ $\ldots,N)$ are the values of the wave function and the dielectric constant at these points. The boundary conditions (\ref{eq:bd1}) and (\ref{eq:bd2}) can be rewritten as
\begin{align}
&\Psi_0 = \frac{2+iq}{2-iq}\Psi_1 +\eta,\label{eq:bd1_a}\\
&\Psi_{N+1} = \frac{2+iq}{2-iq}\Psi_N, \label{eq:bd1_b}
\end{align}
where $q\equiv k\Delta$ is dimenionless. The constant term $\eta\equiv-4iq/(2-iq)$ in Eq.~(\ref{eq:bd1_a}) is due to the incoming wave $\Psi^-_L$ as mentioned previously, and by dropping it we recover the generalized eigenvalue problem that determines the CF states with purely outgoing boundary condition \cite{LiThesis}:
\be
H\bm{\psi}_m = -q_m^2\bm{\e\psi}_m.\label{eq:H_1D}
\ee
$\bm{\psi}_m$ is a column vector containing the values $\Psi_i$ in the $m$th CF state and $\bm{\e}$ is a diagonal matrix with elements $\e_i$. The CF frequencies (similar to the resonances or the poles of the $S$ matrix) is given by $k_m = q_m/\Delta$. Note that the real-valued free-space wave vector $k$, instead of the complex-valued CF frequencies $k_m$'s, appears in the $N\times N$ non-Hermitian Hamiltonian $H$:
\be
H_{ij} = \left[-2 + \frac{2+iq}{2-iq}(\delta_{i,1} + \delta_{i,N})\right]\delta_{ij} +(\delta_{i+1,j}+\delta_{i-1,j}),
\ee
which is tri-diagonal. We note that it is not possible to write the corresponding equation for resonances as such a generalized eigenvalue problem, because $H$ would contain the complex-valued resonance frequencies yet to be determined.

Now with the constant term $\eta$ in the boundary condition (\ref{eq:bd1_a}), the wave function in the scattering problem is given by
\be
H\bm{\Psi} + \bm{F} = -q^2\bm{\e\Psi},\label{eq:H_scat_1D}
\ee
and the inhomogeneous term $\bm{F}$ is a column vector with a single non-zero element, i.e., $F_{1}=\eta$ when light is incident from the left. Similarly, for the scattering of a right-incident wave, the only non-zero element of $\bm{F}$ is $F_{N}=\eta$.
The equation above can be put in a more explicit form to obtain $\bm{\Psi}$:
\be
\bm{\Psi} = -[H+q^2\bm{\e}]^{-1}\bm{F}.\label{eq:inversion}
\ee
The transmission and reflection coefficients can then be calculated using
\be
r_L = \frac{2\Psi_1-(2+iq)}{2-iq},\; t_L = \frac{2\Psi_N}{2-iq}\label{eq:rt1D}
\ee
for left incidence (and similarly for right incidence), and the $S$ matrix is given by
\be
S =
\begin{pmatrix}
r_L & t_L\\
t_R & r_R
\end{pmatrix}
\ee
with $t_L=t_R$ when the system has Lorentz reciprocity \cite{Haus,Collin,Landau}. 

For comparison, here we also discuss the modal expansion approach to construct $S$. By inserting $\bm{\Psi}=\sum_m a_m \bm{\psi}_m$ to Eq.~(\ref{eq:H_scat_1D}) and utilizing Eq.~(\ref{eq:H_1D}), we find
\be
\bm{\Psi} = -\frac{\Delta}{L}\sum_m \frac{\bm{\psi}_m\bm{\psi}_m^T}{q^2-q_m^2}\bm{F},\label{eq:CFexp}
\ee
where we have used the ``self-orthogonality" of the CF states \cite{Ge_pra10} in the following form:
\be
\bm{\psi}_m^T\bm{\e}\bm{\psi}_n = \frac{L}{\Delta}\delta_{m,n}.
\ee
For left incidence and taking $\Delta\to0$, we find
\be
\psi(x) = \frac{-\eta}{\Delta L}\sum_m \frac{\psi_m(x)\psi_m(0)}{k^2-k^2_m} = \frac{2ik}{L}\sum_m \frac{\psi_m(x)\psi_m(0)}{k^2-k^2_m}.\label{eq:psi_expansion}
\ee
This expression is identical to that used in the standard derivation of the $\cal R$-matrix method in the CF basis, which we outline below.

As mentioned in the introduction, the expansion of $\psi$ in a finite number of basis functions introduces an artifact to the normal derivative at the LSS. Therefore, the standard derivation of the $\cal R$-matrix method resorts to the Green's theorem instead to take into consideration the boundary condition of $\psi$, resulting in
\be
a_m = \frac{1}{L}\int_0^L \e\psi_m\Psi dx = \frac{1}{L}\frac{\left[\Psi\partial_x\psi_m - \psi_m\partial_x\Psi\right]^L_0}{k^2-k_m^2}. \label{eq:am_Rmatrix}
\ee
We note that the scattered waves in $\Psi$, as well as the CF states, produce $-ik$ ($ik$) at $x=0$ ($L$) after taking the derivative, and the corresponding boundary terms above are all canceled. Hence we find $\left[\Psi\partial_x\psi_m - \psi_m\partial_x\Psi\right]^L_0 = -\left[ \Psi_L^-(0) \partial_x\psi_m(0) - \psi_m(0)\partial_x\Psi_L^-(0)\right] = 2ik\psi_m(0)$, with which we immediately recover Eq.~(\ref{eq:psi_expansion}). Once $\bm{\psi}_m$'s and $q_m$'s are known, the $S$ matrix can be constructed again by applying Eq.~(\ref{eq:rt1D}) and the corresponding expressions for $r_R,t_R$.

\begin{figure}[t]
\begin{center}
\includegraphics[width=\linewidth]{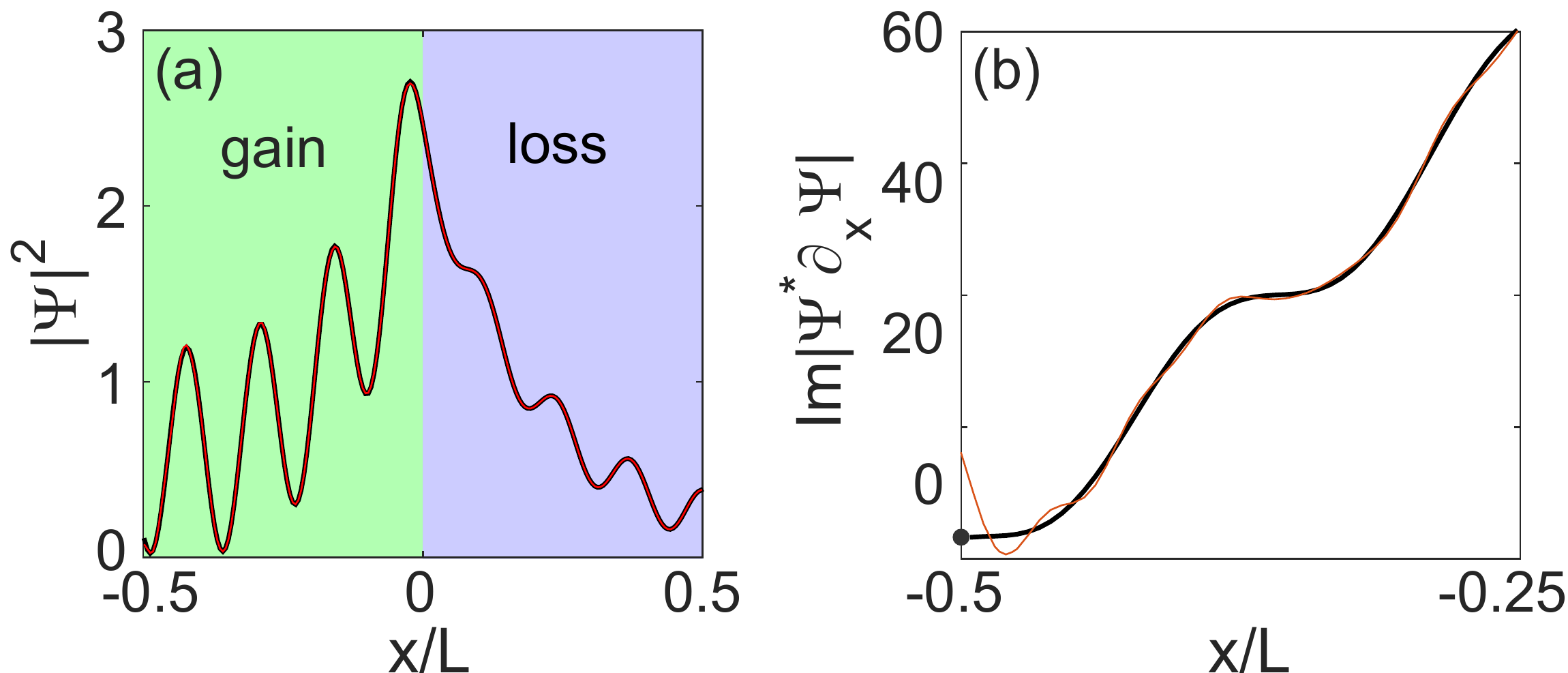}
\caption{Total wave function (a) and its flux (b) depicted by black thick lines for a half-gain-half-loss microcavity with light incident from the left. The wave vector $k=12/L$ and refractive indices $n_1=n_2^*=2 - 0.2i$ are used. The expansion (\ref{eq:CFexp}) with 50 CF states is plotted by the red thin lines as a comparison, which can barely be distinguished from the black line in (a) but shows a significant deviation near the left boundary in (b). The black dot in (b) shows the analytical result at $x=-L/2$ given by Eq.~(\ref{eq:flux1D}).}
\label{fig:1D}
\end{center}
\end{figure}

In Fig.~\ref{fig:1D}(a) we show the total wave function inside a half-gain-half-loss optical microcavity with $\cal PT$ symmetry \cite{Bender1,Bender2,Bender3} and a left incident wave, whose refractive index satisfies $n(-x)=n^*(x)$ \cite{El-Ganainy_OL07,Moiseyev,Musslimani_prl08,Makris_prl08,Guo,Longhi,CPALaser,Robin,EP9,Ruter,Lin,Feng_NM,Feng2,Walk,Hodaei,Yang,Chang,conser2D}.
Good agreement between $\Psi$'s given by Eqs.~(\ref{eq:inversion}) and (\ref{eq:CFexp}) are obtained using 50 CF states. Nevertheless, the artifact of $\partial_x\Psi$ at the boundary of the microcavity in the $\cal R$-matrix approach can be readily seen in Fig.~\ref{fig:1D}(b), where we plot the optical flux given by $\im{\Psi^*\partial_x\Psi}$ (up to a prefactor). 
The BVP approach, on the other hand, gives a good agreement with the analytical result \cite{reciprocity}
\be
\im{\Psi^*\partial_x\Psi}_L = k(1-|r_L|^2),\;r_L = \frac{{\cal G} + i{\cal F}}{\cal D},\label{eq:flux1D}
\ee
where ${\cal G}\equiv qs_1s_2$, ${\cal F}\equiv u_1s_1c_2 + u_2s_2c_1$, ${\cal D}\equiv c_1c_2 - gs_1s_2 - i(h_1s_1c_2 + h_2s_2c_1)$, and $q=(n_1/n_2-n_2/n_1)/2$, $s_j=\sin(n_jk L/2)$, $c_j=\cos(n_j k L/2)$, $u_j = (n_j-1/n_j)/2$, $g=(n_1/n_2+n_2/n_1)/2$, $h_j = (n_j+1/n_j)/2\;(j=1,2)$.
\section{Nonlocal BVP in 2D}

In this section we elucidate how the $S$ matrix is constructed in our scheme as a nonlocal BVP in 2D. Similar to the 1D case discussed in the previous section, we show that the non-Hermitian Hamiltonian $H$ in our approach is also the one determines the CF states with purely outgoing boundary condition.

\subsection{Construction of the $S$ matrix}
\label{sec:main}
Before we proceed, we note that in the calculation of a CF state or a resonance, one basically assumes a homogenous source residing inside an optical microcavity, reflected by the imaginary part of its complex frequency. In a scattering problem, the source is an external one instead and one needs to find a way to distinguish in the boundary condition the known incident wave and the unknown scattered waves. In the 1D case shown in the previous section, the incident wave adds an inhomogeneous term (i.e., $\bm{F}$) to the non-Hermitian eigenvalue problem that determines the outgoing CF states [c.f. Eqs.(\ref{eq:H_1D}) and (\ref{eq:H_scat_1D})]. This separation of incident and outgoing waves holds even when the boundary condition becomes nonlocal in 2D as we show below.

\begin{figure}[t]
\begin{center}
\includegraphics[width=0.8\linewidth]{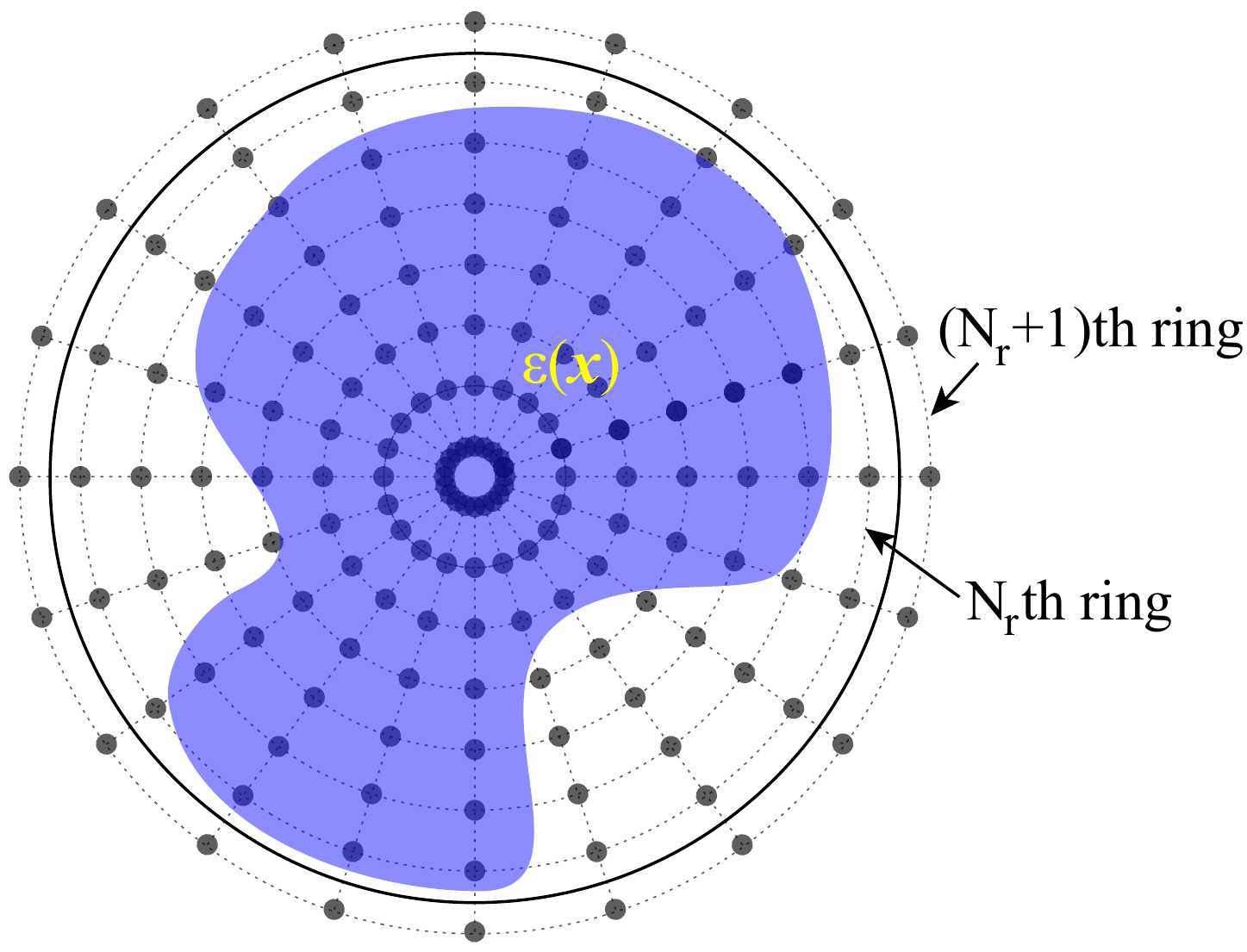}
\caption{Schematic of an optical microcavity (shaded area) and the circular LSS (solid line) in 2D. The finite-difference grid is indicated by the dots and dashed lines.}
\label{fig:schematic}
\end{center}
\end{figure}

To illustrate this property, we again consider the scalar Helmholtz equation and use the polar coordinates. A circular LSS of radius $R$ encloses the optical microcavity with dielectric constant $\e(\bx)$ (see Fig.~\ref{fig:schematic}), outside which we assume $\e(\bx)=n_e^2>1$ and adopt
 \be
 \Psi^-_m = \frac{H^-_m(n_ekr)}{H^-_m(n_ekR)}e^{im\theta}, \; \Psi^+_m = \frac{H^+_m(n_ek^*r)}{H^+_m(n_ek^*R)}e^{im\theta}\label{eq:channels}
 \ee
as our incoming and outgoing channels. Here $H^\mp_m$ are the second and first Hankel functions and $m$ is the angular momentum number, which also serves as the channel index. Note that we do not restrict the free-space wave vector $k$ to be real, which enables the calculation of the resonances as the complex-valued poles of the $S$ matrix. 

Suppose the incident wave impinges on the LSS in the $m_0$th channel $\Psi^-_{m_0}$. The total field and its radial derivative outside the LSS can then be written as
\begin{align}
&\Psi_>  = \Psi^-_{m_0} + \sum_m S_{m,m_0} \Psi^+_m, \label{eq:phi_continuity}\\
&\frac{\partial \Psi_>}{\partial r}  = V^-_{m_0}(r)\,e^{im_0\theta} + \sum_m S_{m,m_0} V^{+*}_m(r)\,e^{im\theta}, \label{eq:dphidr_continuity}
\end{align}
where $V^\pm_{m}(r) \equiv n_ekH^{\pm'}_m(n_ekr)/H^\pm_m(n_ekR)$. Next we discretize the 2D space on a polar grid, with the circular LSS placed at the middle of the $N_r$th and $(N_r+1)$th rings (see Fig.~\ref{fig:schematic}). On each ring there are $N_\theta$ equally spaced grid points with spacing $\Delta_\theta=2\pi/N_\theta$. We then write the total field on the $N_r$th and $(N_r+1)$th rings as
\begin{align}
&\Psi_{N_r,\,\nu}  = \sum_m b_m \,e^{im\theta_\nu}, \label{eq:ringN}\\
&\Psi_{N_r+1,\,\nu}  = \sum_m b_m (1 + c_m\Delta_r)\,e^{im\theta_\nu}, \label{eq:ringN+1}
\end{align}
where $\nu$ is the index for the grid points in the azimuthal direction, $\theta_\nu$ is the coresponding azimuthal angle, and $\Delta_r$ is the uniform spacing between two consecutive grid points in the radial direction.

To derive the nonlocal boundary condition and the $S$ matrix, we eliminate the two coefficients $b_m,c_m$ in the example of TM polarization (with the electric field perpendicular to the 2D cavity plane), and the case of TE polarization can be treated in a very similar fashion. Using the continuity of both $\Psi$ and its radial derivative, the left hand sides of Eqs.~(\ref{eq:phi_continuity}) and (\ref{eq:dphidr_continuity}) on the LSS can be approximated by
\begin{align}
\frac{\Psi_{N_r+1,\,\nu} + \Psi_{N_r,\,\nu}}{2} &= \sum_m b_m (1 + \frac{c_m\Delta_r}{2})\,e^{im\theta_\nu}, \label{eq:discretizeCFBC2d-0}\\
\frac{\Psi_{N_r+1,\,\nu} - \Psi_{N_r,\,\nu}}{\Delta_r} &= \sum_m b_m\, c_m\,e^{im\theta_\nu}, \label{eq:discretizeCFBC2d-1}
\end{align}
which lead to
\begin{align}
S_{m,m_0} &= \frac{b_m c_m - V^-_{m_0}\delta_{m,m_0}}{V^{+*}_m}, \label{eq:S0}\\
S_{m,m_0} &=  b_m \left(1+c_m\frac{\Delta_r}{2}\right) - \delta_{m,m_0},\label{eq:S0a}
\end{align}
and we have dropped the argument $R$ in $V^\pm_m$. The product $b_mc_m$ can be eliminated to derive a more concise form of the $S$ matrix:
\be
S_{m,m_0} = \frac{b_m-(1-V^-_{m_0}\frac{\Delta_r}{2})\delta_{m,m_0}}{1-V^{+*}_m\frac{\Delta_r}{2}}.\label{eq:S}
\ee
This is the expression we will use in our numerical examples, which requires obtaining $b_m$ using the Fourier transform of $\Psi_{N_r,\nu}$'s:
\be
b_m = \sum_{\nu} \frac{\Delta_\theta}{2\pi} e^{-im\theta_{\nu}}\Psi_{N_r,\,\nu}.\label{eq:bm}
\ee
To find $\Psi_{N_r,\nu}$, we eliminate $S_{m,m_0}$ in Eqs.~(\ref{eq:S0}) and (\ref{eq:S0a}) and derive an expression for $b_mc_m$, which when
substituted into Eq.~(\ref{eq:ringN+1}) gives our nonlocal scattering boundary condition:
\be
\Psi_{N_r+1,\,\nu} = \sum_{\nu'} O_{\nu,\,\nu'}\Psi_{N_r,\,\nu'} + f^{(m_0)}_\nu, \label{eq:discretizeCFBC2d}
\ee
where
\bea
O_{\nu,\,\nu'} &\equiv& \frac{\Delta_\theta}{2\pi}\sum_{m}\frac{H^{+}_m(n_ek^* R^+)}{H^+_m(n_ek^*R^-)} e^{im(\theta_\nu-\theta_{\nu'})}, \\
f^{(m_0)}_\nu &=& \frac{(V_{m_0}^- - V_{m_0}^{+*})\Delta_r}{1 - V_{m_0}^{+*}\frac{\Delta_r}{2}} e^{im_0\theta_\nu},
\eea
and $R^-$ is the radius of the $N_r$th ring.
Note that the term $f^{(m_0)}_\nu$ in Eq.~(\ref{eq:discretizeCFBC2d}) is the manifestation of the incoming wave, and by dropping it  we again recover the (nonlocal) outgoing boundary condition that defines the CF states \cite{science}, similar to Eq.~(\ref{eq:bd1_a}) in 1D.

Eq.~(\ref{eq:discretizeCFBC2d}) can then be inserted into the discretized Helmholtz equation on the $N$th ring
to eliminate $\Psi_{N_r+1,\,\nu}$ \cite{LiThesis}, which gives rise to the following matrix equation:
\be
(H + k^2 \bm{\e}) \bm{\psi} + \bm{F}^{(m_0)} = 0.\label{eq:H_2D}
\ee
The column vector $\bm{\psi}\equiv \sqrt{r}\bm{\Psi}$ represents the total wave function, i.e.,
\be
\bm{\psi} = (\psi_{1,1}\ldots\psi_{1,N_\theta}\,\psi_{2,1}\ldots\psi_{N_r,1}\ldots\psi_{N_r,N_\theta})^T.
\ee
and $\bm{F}^{(m_0)}$ is a column vector of zeros except for the last $N_\theta$ elements, which are given by $Rf^{(m_0)}_\nu/(\sqrt{R^-}\Delta_r^2)\, (\nu=1,2,\ldots,N_\theta)$. $\bm{\e}$ has the same form as its 1D counterpart, i.e., a diagonal matrix with the values of the dielectric constant on the discretized grid.
$H$ has $(N_r\times N_\theta)$ rows and columns; it is the same non-Hermitian Hamiltonian that determines the CF states \cite{science,LiThesis}:
\be
H\bm{\psi}_n =- k_n^2 \bm{\e} \bm{\psi}_n,\label{eq:HCF_2D}
\ee
and it consists a banded matrix $H_0$ and a $(N_\theta\times N_\theta)$ block $H'$ in the lower right corner. $H_0$ is symmetric with nonzero elements on the 0, $\pm1$, and $\pm N_\theta$ diagonals
\be
\left\{
\ba{lcl}
H_{(\mu-1)N_\theta+\nu,\,(\mu-1)N_\theta+\nu} &=& -\frac{2}{(\Delta_r)^2} -\frac{2}{(r_\mu\Delta_\theta)^2}, \\
H_{(\mu-1)N_\theta+\nu,\,(\mu-1)N_\theta+\nu+1} &=& \frac{1}{(r_\mu\Delta_\theta)^2}, \\
H_{(\mu-1)N_\theta+\nu,\,\mu N_\theta+\nu} &=& \frac{r_{\mu+\frac{1}{2}}}{(\Delta_r)^2\sqrt{r_\mu r_{\mu+1}}},
\ea\right.\nonumber
\ee
and $H'$ comes from the nonlocal boundary condition (\ref{eq:discretizeCFBC2d}):
\be
H'_{\nu,\,\nu'} = \frac{1}{(\Delta_r)^2}\frac{R}{R^-}O_{\nu,\,\nu'}.
\ee
$H'$ can be checked to be also symmetric using $H^+_{-m}(z) = (-1)^m H^+_m(z)$ and the definition of $O_{\nu,\,\nu'}$ in Eq.~(\ref{eq:discretizeCFBC2d}).
Once $\bm{\psi}$ is obtained for each incoming channel by solving Eq.~(\ref{eq:H_2D}), i.e.,
\be
\bm{\psi} = -(H + k^2 \bm{\e})^{-1}\bm{F}^{(m_0)} \label{eq:inversion_2D}
\ee
as in the 1D case, we know immediately the Fourier coefficients $\{b_m\}$ from Eq.~(\ref{eq:bm}), with which the construction of the $S$ matrix is completed using Eq.~(\ref{eq:S}).

In the limit of a fine grid ($\Delta_r\to0$), Eq.~(\ref{eq:S}) becomes
\be
S_{m,m_0} \xrightarrow[]{\Delta_r\to 0} b_m - \delta_{m,m_0}.\label{eq:S_asymptotic}
\ee
We note that the second term in this expression does not depend on the dielectric constant inside the LSS, and hence it can be regarded as the result of a ``direct scattering" process \cite{Fan}. The first term then corresponds to the ``resonance-assisted" scattering process, and to understand its determining factors, we resort to the modal expansion of $\bm\psi$ in the CF basis, which takes the following form in 2D:
\be
\bm{\psi} = -\frac{\Delta_r\Delta_\theta}{\pi R^2}\sum_n \frac{\bm{\psi}_n\bm{\psi}_n^T}{k^2-k_n^2} \bm{F}^{(m_0)}.\label{eq:CFexp_2D}
\ee
Note that we have used the following normalization of the CF basis
\be
{\bm\psi}_n^T{\bm\e}{\bm\psi}_{n'} = \frac{\pi R^2}{\Delta_r\Delta_\theta}\,\delta_{n,n'}
\ee
such that ${\bm\Psi}_n\equiv {\bm{\psi}_n}/{\sqrt{r}}$ is dimensionless and its value does not scale with the discretization, i.e., the expression above becomes
\be
\int_\text{system}\e\Psi_n\Psi_{n'} r\,dr\,d\phi = \pi R^2\delta_{n,n'}
\ee
in the continuous limit.

$\bm{\Psi}_n$ given by Eq.~(\ref{eq:CFexp_2D}) has the typical resonant denominator with $k_m$'s being the CF frequencies. If we project each CF state at the LSS onto the outgoing channel function $\Psi^+_m$ (which is equivalent to a Fourier transform), i.e.,
\be
\bm{\Psi}_n|_{r=R} = \sum_m z^{(n)}_m \Psi^+_m|_{r=R} = \sum_m z^{(n)}_m e^{im\theta},\label{eq:channelExpasion}
\ee
the inner product $\bm{\psi}_n^T\bm{F}^{(m_0)}$ singles out the $m_0$th Fourier coefficient $z_{m_0}$:
\be
\bm{\psi}_n^T\bm{F}^{(m_0)} = \frac{2 R}{\Delta_r\Delta_\theta}(V_{m_0}^- - V_{m_0}^{+*})z^{(n)}_{m_0},
\ee
and consequently,
\be
S_{m,m_0} \xrightarrow[]{\Delta_r\to 0} \frac{2}{R}(V_{m_0}^{+*} - V_{m_0}^- ) {\cal R}_{m,m_0} - \delta_{m,m_0},\label{eq:S_Rmatrix}
\ee
where
\be
{\cal R}_{m,m_0} = \sum_n \frac{z^{(n)}_{m}z^{(n)}_{m_0}}{k^2-k_n^2}\label{eq:Rmatrix}
\ee
is the $\cal R$ matrix \cite{Lane}. This expression indicates clearly that the contribution of a particular CF state to the scattering process is not only dependent on the resonant denominator; it also depends on the spatial overlaps between this CF state and the incoming and outgoing channels at the LSS, represented by $z^{(n)}_{m_0}$ and $z^{(n)}_{m}$ respectively. In the simplest example where the system is isotropic and the angular momentum is a good quantum number (e.g., a circular microdisk cavity), only the CF states with the same angular momentum as the incoming (and outgoing) channel contribute to the scattering process.

To show that $S_{m,m_0}$ given by Eq.~(\ref{eq:S_Rmatrix}) is consistent with the standard $\cal R$-matrix result in the CF basis, we apply the Green's theorem to the interior of the LSS, which gives us
\begin{align}
a_n &= \frac{1}{\pi R^2}\int_\text{system} \e\Psi_n\Psi d\bm{x} \nonumber \\
&= \frac{1}{\pi R}\frac{\int_\text{LSS}\left[\Psi\partial_r\Psi_n - \Psi_n\partial_r\Psi\right]_{R} d\theta}{k^2-k_n^2}. \label{eq:am_Rmatrix_2D}
\end{align}
Similar to the 1D case, the outgoing channels are cancelled in the boundary integral, which can be proved rigorously by applying the Green's theorem again to the \textit{exterior} of the system, where both $\Psi$ and $\psi_n$ satisfy
\be
[\nabla^2 + n_e^2 k^2]X = 0,\quad X=\Psi,\Psi_n.
\ee
Note that the same wave vector $k$ for the total field $\Psi$ and the CF states $\psi_n$ is crucial to remove the outgoing waves from the boundary integral in Eq.~(\ref{eq:am_Rmatrix_2D}), and we end up with
\be
a_n = \frac{1}{\pi R}\frac{\int_\text{LSS}\left[\Psi^-_{m_0}\partial_r\Psi_n - \Psi_n\partial_r\Psi^-_{m_0}\right]_{R} d\theta}{k^2-k_n^2}
\ee
for incoming wave in the $m_0$th channel. Unlike the $\cal R$ matrix in other basis \cite{Lane}, here the $S$ matrix does not appear in the expansion coefficient $a_n$.
By realizing that the expansion (\ref{eq:channelExpasion}) holds not just on the LSS but also in the exterior region, we find
\be
\left.\partial_r\Psi_n\right|_{r=R} =  \sum_m b_m V_m^{+*}e^{im\theta}
\ee
and the total wave function in the interior of the system is given by
\be
\Psi(\bx) = \frac{2}{R}(V_{m_0}^{+*} - V_{m_0}^- )\sum_n \frac{z^{(n)}_{m_0}\Psi_n(\bm{x})}{k^2-k_n^2}.
\ee
Once we substitute the $\Psi(\bx)$ on the LSS by Eq.~(\ref{eq:phi_continuity}) and project both sides of the equation above onto the outgoing channels, we immediately recover the $S$ matrix given by Eq.~(\ref{eq:S_Rmatrix}).

\subsection{Results}

\begin{figure}[t]
\begin{center}
\includegraphics[width=\linewidth]{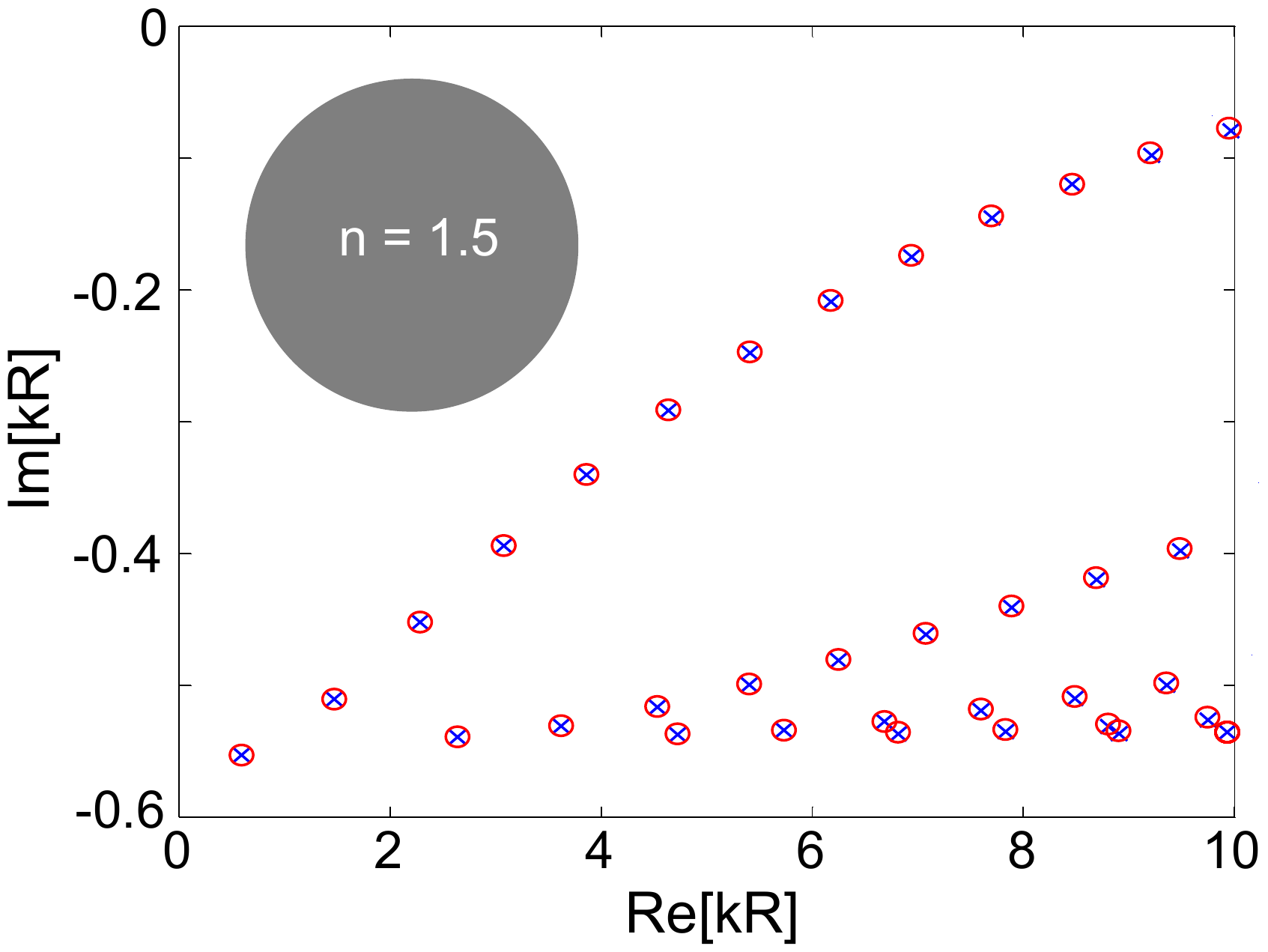}
\caption{Resonances of a microdisk cavity with a uniform index $n=1.5$. The crosses are the analytical results given by Eq.~(\ref{eq:qb}) and the circles are the poles of the $S$ matrix constructed using Eq.~(\ref{eq:S}).}
\label{fig:qb}
\end{center}
\end{figure}

To test our approach based on a nonlocal BVP in 2D, we first calculate the poles of the $S$ matrix for a circular microdisk cavity with a uniform index. As mentioned in the introduction, the poles of the $S$ matrix correspond to complex-valued resonances of the optical microcavity. This connection is due to the diverging eigenvalues of the $S$ matrix at its poles, meaning that an infinitesimal incoming amplitude leads to finite outgoing waves. For a circular microdisk cavity with a uniform index, the angular momentum number $m$ is a conserved quantity and the LSS is chosen as the disk boundary. The TM resonances can be found by solving the following analytical expression:
\be
\frac{H_m^{+'}(kR)J_m^{+'}(nkR)}{H_m^{+}(kR)J'_m(nkR)}=n.\label{eq:qb}
\ee
In Fig.~\ref{fig:qb} we compare the poles of the $S$ matrix calculated by the BVP approach and this analytical expression, and good agreement is found for different poles with $m\in[0,12]$.

\begin{figure}[t]
\begin{center}
\includegraphics[width=\linewidth]{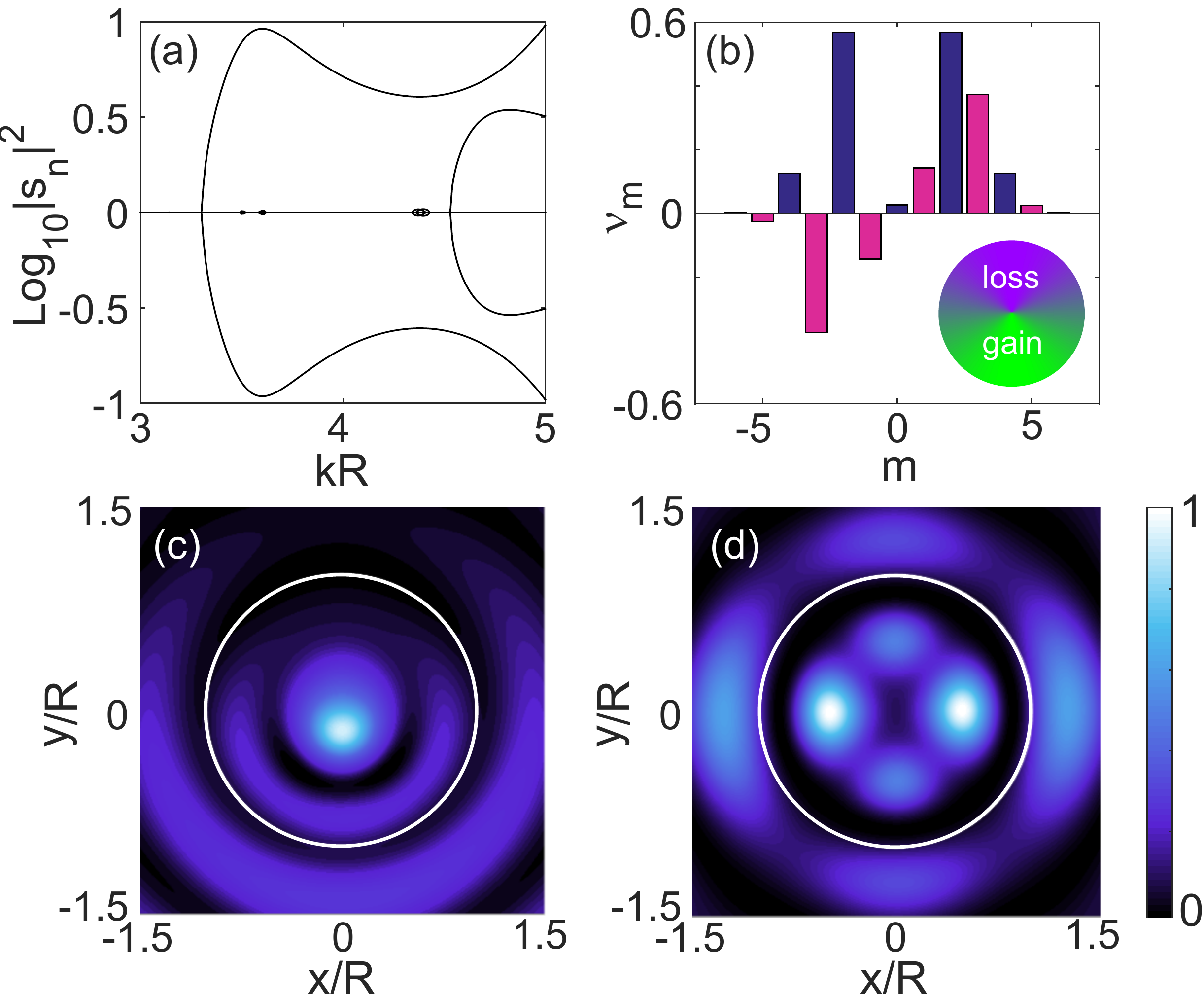}
\caption{(a) Spontaneous symmetry breaking of $S$-matrix eigenvalues $s_n$ in a microdisk cavity with $\cal PT$ and $\cal RT$ symmetries. Its refractive index is given by $n(\bm{x})=1.5+0.4\sin\theta$, the imaginary part of which is shown schematically by the inset in (b). (b) A real-valued eigenvector of $S$ at $kR=4$ in the $\cal PT$- and $\cal RT$-symmetric phase. The blue (pink) bars show symmetric and antisymmetric components with opposite $m$'s. The corresponding wave function is shown in (d), where the cavity boundary is marked by the white circle. The wave function of a scattering eigenstate in the broken-symmetry phase at $kR=4$ is shown in (c) as a comparison.}
\label{fig:lambda}
\end{center}
\end{figure}

Next we inspect the $S$ matrix constructed by the BVP approach from a different perspective, i.e., the symmetry property of its eigenvalues in the presence of $\cal PT$ and $\cal RT$ symmetries. It was found that in a $\cal PT$-symmetric system $s_m$'s undergo spontaneous symmetry breaking as a function of frequency or system size \cite{CPALaser}: $\cal PT$ symmetry warrants $s_m^*=s_{m'}^{-1}$. When $m,m'$ in this expression are the same, the corresponding eigenstate of the $S$ matrix is in the $\cal PT$-symmetric phase with $|s_m|=1$, i.e., the optical flux in the corresponding scattering eigenstate is conserved, even though the system is non-Hermitian in the presence of gain and loss. In the $\cal PT$-broken phase $m\neq m'$, and we find $|s_m|=|s_{m'}|^{-1}$ instead (or equivalently, $\log_{10}|s_m|^2=-\log_{10}|s_{m'}|^2$), which represent a pair of amplified and attenuated scattering eigenstates.

In Fig.~\ref{fig:lambda} we show the eigenvalues of a microdisk cavity with refractive index $n(\bm{x})=1.5+0.4\sin\theta$, which satisfies not only $\cal PT$ symmetry (here $\cal P$ changes $\theta$ to $-\theta$) but also $\cal RT$ symmetry \cite{EP9,conser2D}, i.e., $n(r,\theta) = n^*(r,\theta+\pi)$. The eigenvalues of the $S$ matrix for an $\cal RT$-symmetric structure display similar spontaneous breaking as those in their $\cal PT$-symmetric counterparts, and these properties are manifested nicely by the $S$ matrix given by Eqs.~(\ref{eq:S}) using the BVP approach [Fig.~\ref{fig:lambda}(a)]. 

It can be easily seen that when a system has both $\cal PT$ and $\cal RT$ symmetries, their symmetric (broken) phases for the $S$ matrix coincide due to the unimodular property of $s_m$. Therefore, the scattering eigenstates in the symmetric phase should also possess simultaneously the properties of both $\cal PT$ and $\cal RT$ symmetries. To understand and differentiate these properties, we turn to the eigenvectors of the $S$ matrix, which are the projection coefficients of the scattering eigenstates onto the incoming and outgoing channels. The cylindrical channels specified by Eq.~(\ref{eq:channels}) are $\cal PT$-symmetric, i.e., ${\cal PT}\Psi^-_m = \Psi^+_m$, where the minus sign introduced by the parity operator (again $\theta\to -\theta$) in the exponent is canceled by performing a time reversal (i.e., $i$ becomes $-i$ in the exponent; it also changes $H^\pm_m$ to $H^\mp_m$). Now if $\Psi^-(r,\theta) = \sum_m v_m \Psi^-_m(r,\theta)$ is the incident wave in an eigenstate of $S$ with eigenvalue $s_n$, then by performing a combined $\cal PT$ operation on its scattered wave (i.e., $\Psi^+(r,\theta) = s_n\sum_m v_m \Psi^+_m(r,\theta)$) the new incoming state $\bar{\Psi}^-(r,\theta)\equiv {\cal PT}\Psi^+(r,\theta) = s_n^*\sum_m v_m^* \Psi^-_m(r,\theta)$ should also be an eigenstate of the $S$ matrix due to $\cal PT$ symmetry. It then follows that $v_m=g v_m^*$ should hold for all $m$'s in the $\cal PT$-symmetric phase, and the proportional constant $g$ can be set as 1 by choosing a proper global phase of $v_m$. In other words, all $v_m$'s can be made real in the $\cal PT$-symmetric phase. If we apply a similar analysis of $\cal RT$ symmetry, we find that the channel functions transform according to ${\cal RT}\Psi^-_m = (-)^m\Psi^+_{-m}$, which leads to $v_{-m}^* = \pm(-)^m v_m$. Therefore, we find
\be
v_{-m} = \pm(-)^m v_m \;(v_m\in\mathbb{R})\label{eq:ves_PTRT}
\ee
as a result of these two symmetries, which is captured nicely by the result of the BVP approach [Fig.~\ref{fig:lambda}(b)].

We also note that a 2D structure with both $\cal PT$ and $\cal RT$ symmetries also has mirror symmetry about the $\theta=\pm\pi/2$ axis (i.e., the axis perpendicular to the parity axis in $\cal PT$ symmetry) \cite{EP9}, which imposes the following property:
\be
v_{-m} = \pm(-)^m v_m \;(v_m\in\mathbb{C}).
\ee
The overall $\pm$ sign corresponds to scattering eigenstates that are even and odd functions about the $\theta=\pm\pi/2$ axis, respectively. Again the mirror symmetry about the $\theta=\pm\pi/2$ axis is observed nicely in the BVP approach, both in the symmetry-broken phase [Fig.~\ref{fig:lambda}(c)] and symmetric phase [Fig.~\ref{fig:lambda}(d)]. In the former ${\cal PT}\Psi\neq\Psi, {\cal RT}\Psi\neq\Psi$ and hence $|\Psi(r,\theta)|\neq|\Psi(r,-\theta)|, |\Psi(r,\theta)|\neq|\Psi(r,\theta+\pi)|$; In the latter we find $|\Psi|$ is symmetric about both the horizontal and vertical axes instead.

\begin{figure}[b]
\begin{center}
\includegraphics[width=\linewidth]{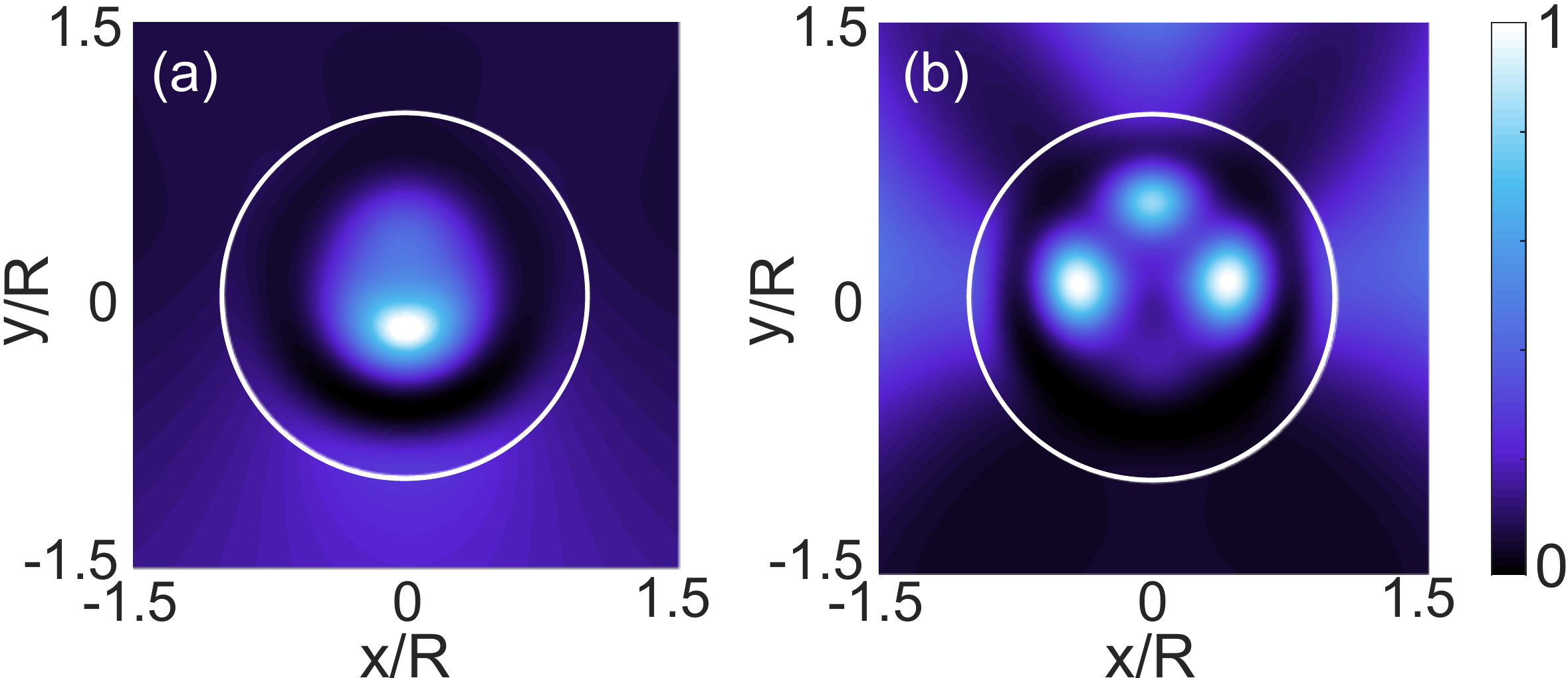}
\caption{Resonant modes corresponding to the scattering eigenstates in Figs.~\ref{fig:lambda}(c) and (d).}
\label{fig:qbPT}
\end{center}
\end{figure}

The corresponding resonant modes with resonant frequencies $k_mR = 3.5042 - 0.2492i,\,4.5989 - 0.6445i$ are shown in Fig.~\ref{fig:qbPT}, which are calculated as the poles of the $S$ matrix constructed using the BVP method. These poles differ by less than $0.01\%$ from the results of an iterative method we outline below, in both their real and imaginary parts. As briefly mentioned in Sec.~\ref{sec:1D}, the difference between a CF state and a resonance lies in the wave vector in the exterior of an optical microcavity: A CF state features a real-valued $k$ while a resonance has the same complex-valued resonant frequency $k_m$ as in the interior of the microcavity. Therefore, if we replace $k$ by a CF frequency $k_m$ in the eigenvalue problem (\ref{eq:HCF_2D}) that determines the CF states and repeat this procedure until $k_m$ converges, we end up with the same complex-valued frequency $k_m$ in both the interior and exterior of the microcavity, which is a resonance.

It is important to note that while the resonant modes of the system have mirror symmetry about the $\theta=\pm\pi/2$ axis here, they do not possess a $\cal PT$- and $\cal RT$-symmetric phase: performing a combined $\cal PT$ or $\cal RT$ operation does not leave a resonant mode unchanged. Its outgoing waves outside the microcavity are now turned into incoming waves, which by definition give a zero of the $S$ matrix, whose complex-valued frequency is the complex conjugate of the original resonance \cite{CPALaser}. Nevertheless, since the resonant modes are the scattering eigenstates at the poles of the $S$ matrix, they bare resemblance to the latter when $S$ is evaluated at a real-valued frequency, as can be seen by comparing Figs.~\ref{fig:lambda}(c),(d) and Figs.~\ref{fig:qbPT}(a),(b).


\section{Conclusion and discussion}

In summary, we have presented a fintie-difference scheme to construct the $S$ matrix for optical microcavities as a BVP. The boundary condition for the total field is simple in 1D but becomes nonlocal in 2D, which appears as an inhomogeneous term and also in the non-Hermitian Hamiltonian that  determines the CF states. We have verified that our approach is consistent with the $\cal R$-matrix method in the basis of CF states, and it addresses the artifact in the normal derivative of the total field typically found in the $\cal R$-matrix approach.

For applications such as enhancing light-matter interactions and sensing, often it requires accurate knowledge of wave function inside and on the boundary of optical microcavities. In such cases, the BVP method proposed here provides an economic alternative to the modal expansion approach, as the latter requires a large number of basis functions to provide the same level of accuracy. For example, at least 500 basis functions and five times more computational time are needed to capture the symmetry properties of the scattering eigenvalues shown in Fig.~\ref{fig:lambda}(a), whether CF states or the orthogonal states with a vanishing radial derivative at the LSS are used \cite{CPA,CPALaser}.

We also note that there are several other efficient numerical approaches to construct the $S$ matrix, such as finite-different-time-domain methods \cite{YangP,Shibata} already mentioned in the introduction and the method of auxiliary sources \cite{Waterman,Maystre}. The advantages of our approach are that it provides a conceptually clear construction and a numerically straightforward implementation, and it can be applied to three dimensional structures using techniques similar to those developed for binary gratings \cite{Moharam}. Our approach can also be applied to a network of optical microcavities \cite{zeromodeLaser}, and it can treat continuous variations of the refractive index both inside and between these cavities, all enclosed by the LSS.

Finally, we note that while the poles of the $S$ matrix are independent of the choice of the incoming and outgoing channels, the eigenvalues of $S$ do depend on such choices in general. Only when the incoming and outgoing channels are transformed in the same way do the eigenvalues of $S$ stay unchanged, because then $S$ merely experiences a similar transformation. In our discussion of 2D TM waves, the specific forms of the channels have been chosen to simplify the notations in the derivation of the nonlocal boundary condition and the $S$ matrix. When different channels are used, for example, by changing the angular dependence of $\Psi_m^-$ to $e^{-im\theta}$ \cite{CPALaser}, we effectively perform a permutation on the incoming channels, which is not a similar transformation with unchanged outgoing channels. Therefore, the $S$-matrix eigenvalues and their symmetric (symmetry-broken) phases change as a result in general. Exploring this freedom of choosing the channel functions may lead to a close resemblance between the spontaneous symmetry breaking of the $S$ matrix and the corresponding close-cavity modes in $\cal PT$- and $\cal RT$-symmetric systems, similar to the finding in 1D heterostructures \cite{conservation}.\\

\noindent \textbf{Funding.} National Science Foundation (NSF) (DMR-1506987). \\

\noindent \textbf{Acknowledgement.} The author acknowledges support from NSF.


\begin{thebibliography}{99}
\bibitem{RKChang} R. K. Chang and A. J. Campillo, \textit{Optical processes in microcavities}
(World Scientific, 1996).
\bibitem{Vahala} K. J. Vahala, \textit{Optical Microcavities} (World Scientific, 2004).
\bibitem{RMP} H. Cao and J. Wiersig, ``Dielectric microcavities: Model systems for wave chaos and non-Hermitian physics,"
Rev. Mod. Phys. \textbf{87}, 61 (2015).
\bibitem{Nockel} J. U. N\"ockel and A. D. Stone. ``Ray and wave chaos in asymmetric resonant optical cavities," Nature (London) \textbf{385}, 45 (1997).
\bibitem{Gmachl} C. Gmachl, F. Capasso, E. E. Narimanov, J. U. N\"ockel, A. D. Stone, J. Faist, D. L. Sivco, and A. Y. Cho. ``High-power directional emission from microlasers with chaotic resonators," Science \textbf{280}, 1556 (1998).
\bibitem{Fan} W. Suh, Z. Wang, and S. Fan, ``Temporal coupled-mode theory and the presence of non-orthogonal modes in lossless multimode cavities,''
    IEEE J. Quant. Elec. \textbf{40}, 1511 (2004).

\bibitem{Wheeler} J. A. Wheeler, ``On the mathematical description of light nuclei by the method of resonating group structure,"
Phys. Rev. \textbf{52}, 1107 (1937).
\bibitem{Wigner} E. P. Wigner and L. Eisenbud, Phys. Rev. \textbf{72}, 29 (1947).

\bibitem{Nagashima} Y. Nagashima. \textit{Scattering Matrix, in Elementary Particle Physics: Quantum Field Theory and Particles}, Vol. 1 (Wiley-VCH, Weinheim, Germany, 2010).
\bibitem{Peskin} M. E. Peskin and D. V. Schroeder, \textit{An Introduction to Quantum Field Theory} (Westview, US, 1995).

\bibitem{Beenakker} C. W. J. Beenakker, ``Random-matrix theory of quantum transport,"
Rev. Mod. Phys. \textbf{69}, 731 (1997).
\bibitem{Newton} R. G. Newton, \textit{Scattering Theory of Wave and Particles} (Dover Publications, Mineola, New York, 2002).
\bibitem{Dicke} R. H. Dicke, ``A computational method applicable to microwave ntetworks,"
J. Appl. Phys. \textbf{18}, 873 (1947).



\bibitem{Gamow} G. Gamow, ``Zur quantentheorie des atomkernes,"
Z. Phys. 51, 204 (1928).

\bibitem{Lane} A. M. Lane and R. G. Thomas, ``R-matrix theory of nuclear reactions,'' Rev. Mod. Phys. \textbf{30}, 257 (1958).
\bibitem{Breit} G. Breit, ``Scattering matrix of radioactive states,"
Phys. Rev. \textbf{58}, 1068 (1940).
\bibitem{Kapur} P. L. Kapur and R. Peierls, ``The dispersion formula for nuclear reactions,"
Proc. Roy. Soc. A \textbf{166}, 277 (1937).
\bibitem{pra06} H.~E.~T\"ureci, A.~D.~Stone and B.~Collier, ``Self-consistent multimode lasing theory for complex or random lasing media," {Phys. Rev. A} {\bf 74}, 043822 (2006).
\bibitem{BlochC} C. Bloch, ``Une formulation unifi\`ee de la th\'eorie des r\'eactions nucl\'eaires,'' Nucl. Phys. \textbf{4}, 503 (1957).
\bibitem{Szmy} R. Szmytkowski and J. Hinze, ``Convergence of the non-relativistic and relativistic R-matrix expansions at the reaction volume boundary," J. Phys. B: At. Mol. Opt. Phys. \textbf{29}, 761 (1996).

\bibitem{YangP} P. Yang and K. N. Liou, ``Finite-difference time domain method for light scattering by small ice crystals in three-dimensional space," J. Opt. Soc. Am. A \textbf{13}, 2072 (1996).
\bibitem{Shibata} T. Shibata and T. Itoh, ``Generalized-scattering-matrix modeling of waveguide circuits using FDTD field simulations,'' IEEE Trans. Microwave Theory Tech. \textbf{46}, 1742 (1998).

\bibitem{BEM} J. Wiersig, ``Boundary element method for resonances in dielectric microcavities,"
J. Opt. A: Pure Appl. Opt. \textbf{5}, 53 (2003).
\bibitem{SCUFFEM} http://homerreid.github.io/scuff-em-documentation/ (accessed July 31, 2017).
\bibitem{Liew} S. F. Liew, L. Ge, B. Redding, G. S. Solomon, and H. Cao, ``Controlling a microdisk laser by local refractive index perturbation,"
    Appl. Phys. Lett. \textbf{108}, 051105 (2016).
\bibitem{Makris_prl08} K.~G.~Makris, R.~El-Ganainy, D.~N.~Christodoulides, and Z.~H.~Musslimani, ``Beam dynamics in $\cal PT$ symmetric optical lattices," Phys. Rev. Lett. {\bf 100}, 103904 (2008).
\bibitem{Bender1} C.~M.~Bender and S.~Boettcher, ``Real spectra in non-Hermitian Hamiltonians naving $\cal PT$ symmetry,"
Phys. Rev. Lett. {\bf 80}, 5243 (1998).
\bibitem{Bender2} C.~M.~Bender, S.~Boettcher, and P.~N.~Meisinger, ``$\cal PT$-symmetric quantum mechanics,"
J. Math. Phys. {\bf 40}, 2201 (1999).
\bibitem{Bender3} C.~M.~Bender, D.~C.~Brody, and H.~F.~Jones, ``Complex extension of quantum mechanics,"
Phys. Rev. Lett. {\bf 89}, 270401 (2002).
\bibitem{El-Ganainy_OL07} R.~El-Ganainy, K.~G.~Makris, D.~N.~Christodoulides, and Z.~H.~Musslimani, ``Theory of coupled optical $\cal PT$-symmetric structures,"
    Opt. Lett. {\bf 32}, 2632 (2007).
\bibitem{Moiseyev} S.~Klaiman, U.~Gunther, and N.~Moiseyev, ``Visualization of branch points in $\cal PT$-symmetric waveguides,"
Phys. Rev. Lett. {\bf 101}, 080402 (2008).
\bibitem{Musslimani_prl08} Z.~H.~Musslimani, K.~G.~Makris, R.~El-Ganainy, D.~N.~Christodoulides, ``Optical solitons in $\cal PT$ periodic potentials," Phys. Rev. Lett. {\bf 100}, 030402 (2008).
\bibitem{Guo} A. Guo, G. J. Salamo, D. Duchesne, R. Morandotti, M. Volatier-Ravat, V. Aimez, G. A. Siviloglou, and D. N. Christodoulides, ``Observation of PT-symmetry breaking in complex optical potentials,"
    Phys. Rev. Lett. \textbf{103}, 093902 (2009).
\bibitem{Ruter} C. E. R\"uter, K. G. Makris, R. El-Ganainy, D. N. Christodoulides, M. Segev, and D. Kip, ``Observation of parity-time symmetry in optics,"     Nature Phys. \textbf{6}, 192 (2010).
\bibitem{Longhi} S. Longhi, ``$\cal PT$-symmetric laser absorber,"
Phys. Rev. A \textbf{82}, 031801(R) (2010).
\bibitem{CPALaser} Y.~D.~Chong, L.~Ge, and A.~D.~Stone, ``$\cal PT$-symmetry breaking and laser-absorber modes in optical scattering systems,"
    Phys. Rev. Lett. {\bf 106}, 093902 (2011).
\bibitem{conservation} L.~Ge, Y.~D.~Chong, and A. D. Stone, ``Conservation relations and anisotropic transmission resonances in one-dimensional $\cal PT$-symmetric photonic heterostructures,"
    Phys. Rev. A {\bf 85}, 023802 (2012).
\bibitem{Robin} P. Ambichl, K. G. Makris, L.Ge, Y. Chong, A. D. Stone, and S. Rotter, ``Breaking of $\cal PT$ symmetry in bounded and unbounded scattering systems," Phys. Rev. X {\bf 3}, 041030 (2013).
\bibitem{EP9} L. Ge and A. D. Stone, ``Parity-time symmetry breaking beyond one dimension: the role of degeneracy,"
Phys. Rev. X \textbf{4}, 031011 (2014).
\bibitem{Lin} Z. Lin, H. Ramezani, T. Eichelkraut, T. Kottos, H. Cao, and D. N. Christodoulides, ``Unidirectional invisibility induced by PT-symmetric periodic structures," Phys. Rev. Lett. \textbf{106}, 213901 (2011).
\bibitem{Walk} A. Regensburger, C. Bersch, M. A. Miri, G. Onishchukov, D. N. Christodoulides, and U. Peschel, ``Parity-time synthetic photonic lattices,"     Nature (London) \textbf{488}, 167 (2012).
\bibitem{Feng_NM} L. Feng, Y.-L. Xu, W. S. Fegadolli, M.-H. Lu, J. E. B. Oliveira, V. R. Almeida, Y.-F. Chen, and A. Scherer, ``Experimental demonstration of a unidirectional reflectionless parity-time metamaterial at optical frequencies,"
    Nature Mater. \textbf{12}, 108 (2013).
\bibitem{Feng2}L. Feng, Z. J.Wong, R.-M.Ma, Y.Wang, and X. Zhang, ``Singlemode laser by parity-time symmetry breaking,"
Science \textbf{346}, 972 (2014).
\bibitem{Yang} B. Peng, S. K. \"Ozdemir, F. Lei, F. Monifi, M. Gianfreda, G. L. Long, S. Fan, F. Nori, C. M. Bender, and L. Yang,
``Parity-time-symmetric whispering-gallery microcavities," Nature Phys. \textbf{10}, 394 (2014).
\bibitem{Chang} L. Chang, X. Jiang, S. Hua, C. Yang, J. Wen, L. Jiang, G. Li, G. Wang, and M. Xiao, ``Parity-time symmetry and variable optical isolation in active-passive-coupled microresonators," Nature Photon. \textbf{8}, 524 (2014).
\bibitem{Hodaei} H. Hodaei, M. A. Miri, M. Heinrich, D. N. Christodoulides, and M. Khajavikhan, ``Parity-time–symmetric microring lasers,"
Science \textbf{346}, 975 (2014).

\bibitem{conser2D} L. Ge, K. G. Makris, D. N. Christodoulides, and L. Feng, ``Scattering in $\cal PT$- and $\cal RT$-symmetric multimode waveguides: Generalized conservation laws and spontaneous symmetry breaking beyond one dimension,''
    Phys. Rev. A \textbf{92}, 062135 (2015).

\bibitem{LiThesis} L. Ge, Ph.~D. thesis, Yale University (2010).

\bibitem{Haus} H. A. Haus, \textit{Waves and Fields in Optoelectronics} (Prentice-Hall, Englewood Cliffs, NJ, 1984), pp. 56–-61.
\bibitem{Collin} R. E. Collin, \textit{Field Theory of Guided Waves} (McGraw-Hill, New York, 1960).
\bibitem{Landau} L. D. Landau and E. M. Lifshitz, \textit{Electrodynamics of Continuous Media} (Pergamon Press, Oxford, 1960).

\bibitem{Ge_pra10}L. Ge, Y. D. Chong, and A. D. Stone, ``Steady-state ab initio laser theory: generalizations and analytic results," Phys. Rev. A {\bf 82}, 063824 (2010).
\bibitem{reciprocity} L. Ge and L. Feng, ``Optical-reciprocity-induced symmetry in photonic heterostructures and its manifestation in scattering $\cal PT$-symmetry breaking,”
    Phys. Rev. A {\bf 94}, 043836 (2016).
\bibitem{science} H.~E.~T\"ureci, L.~Ge, S.~Rotter and A.~D.~Stone, ``Strong interactions in multimode random lasers," Science {\bf 320} 643 (2008).


\bibitem{CPA} Y.~D.~Chong, L.~Ge, H.~Cao, and A.~D.~Stone, ``Coherent perfect absorbers: time-reversed lasers," {Phys. Rev. Lett.} {\bf 105}, 053901 (2010).

\bibitem{Waterman} P. C. Waterman, ``Matrix formulation of electromagnetic scattering," Proc. IEEE \textbf{53}, 805 (1965).
\bibitem{Maystre} D.~Maystre, S.~Enoch and G.~Tayeb, {\it Scattering Matrix Method Applied to Photonic Crystals}, edited by K.~Yasumoto ({CRC} Press, 2010).
\bibitem{Moharam} M. G. Moharam, Eric B. Grann, and Drew A. Pommet, ``Formulation for stable and efficient implementation of the rigorous coupled-wave analysis of binary gratings," J. Opt. Soc. Am. A \textbf{12}, 1068 (1995).
\bibitem{zeromodeLaser} L. Ge, ``Symmetry-protected zero-mode laser with a tunable spatial profile," Phys. Rev. A \textbf{95}, 023812 (2017).
\end{thebibliography}
\end{document}